\documentclass[prl,twocolumn,showpacs,preprintnumbers,amsmath,amssymb,superscriptaddress]{revtex4}

\usepackage{graphicx}
\usepackage{dcolumn}
\usepackage{bm}

\usepackage{amsmath}
\begin{document}

\newcommand{\lw}[1]{\smash{\lower2.ex\hbox{#1}}}

\title{ LDA + Negative $U$ Solves a Puzzle of too Large Calculated Magnetic Moment 
in Iron-based Superconductor LaFeAsO$_{1-x}$F$_{x}$ }

\author{H.~Nakamura}
\email{nakamura.hiroki@jaea.go.jp}
\affiliation{CCSE, Japan Atomic Energy Agency, 6--9--3 Higashi-Ueno,
Taito-ku Tokyo 110--0015, Japan}
\author{N.~Hayashi} 
\email{hayashi.nobuhiko04@jaea.go.jp}
\affiliation{CCSE, Japan Atomic Energy Agency, 6--9--3 Higashi-Ueno,
Taito-ku Tokyo 110--0015, Japan}
\affiliation{CREST (JST), 4--1--8 Honcho, Kawaguchi, Saitama 332--0012,
Japan}
\author{N.~Nakai}
\email{nakai.noriyuki@jaea.go.jp}
\affiliation{CCSE, Japan Atomic Energy Agency, 6--9--3 Higashi-Ueno,
Taito-ku Tokyo 110--0015, Japan}
\affiliation{CREST (JST), 4--1--8 Honcho, Kawaguchi, Saitama 332--0012,
Japan}
\author{M.~Machida}
\email{machida.masahiko@jaea.go.jp}
\affiliation{CCSE, Japan Atomic Energy Agency, 6--9--3 Higashi-Ueno,
Taito-ku Tokyo 110--0015, Japan}
\affiliation{CREST (JST), 4--1--8 Honcho, Kawaguchi, Saitama 332--0012,
Japan}

\date{June 30, 2008}

\begin{abstract} 
  
A puzzle in the iron-based superconductor LaFeAsO$_{1-x}$F$_{x}$ is that the magnetic moment obtained 
by first-principle electronic structure calculations is unexpectedly much larger than 
the experimentally observed one. For example, the 
calculated value is $\sim 2.0 \mu_{\rm B}$ in the mother compound, while it is $\sim 0.3 \mu_{\rm B}$ in experiments.
We find that the puzzle is solved within the framework LDA + $U$ by expanding
the $U$ value into a slightly negative range. We show $U$ dependence of the obtained magnetic moment in 
both the undoped $x=0.0$ and doped $x = 0.125$.
These results reveal that the magnetic moment is drastically reduced 
when entering to the slightly negative range of $U$.
Moreover, the negative $U$ well explains other measurement data, e.g., lattice constants 
and electronic DOS at the Fermi level. 
We discuss possible origins of the negative $U$ in these compounds.

\end{abstract}
%

\maketitle

Very recently, iron based novel superconductors have attracted much attention, since its discovery may
open another pathway toward the room temperature superconductor. 
Consequently, the superconducting critical temperature has 
exceeded 50K within just a few months since the start of the high-$T_{\rm c}$ race \cite{kamihara}.
At the same time, several theoretical mechanisms have been proposed \cite{mechanism}.

Here, we summarize the experimental results accumulated since the discovery.
The typical undoped mother compound LaFeAsO shows an antiferromagnetic spin density wave (SDW) ordering 
after the lattice structural transition 
from the tetragonal to the orthorhombic structure at $\sim$ 150K 
when decreasing the temperature\cite{neutron,nomura}.
The SDW disappears with doping carriers via chemical substitution of a part of O by F, and 
the superconducting phase instead emerges from $x \sim 0.05$ in LaFeAsO$_{1-x}$F$_{x}$. 
Thus, the superconductivity has a close relationship to the magnetism and 
the lattice distortion.
In addition, the optical absorption, photoemission and NMR studies 
reported that a large pseudo-gap 19 meV $\sim$ 100 meV opens even above 
the superconducting transition $T_{\rm c}$\cite{pseudogap,dong}.
These imply an anomalous DOS suppression at the Fermi level by development
of an order or other reasons. However, the clear origin still remains unsolved.

Besides the above measurement activities, the new material discovery 
has stimulated first-principle electronic structure 
calculations \cite{abinitio_matome,ishibashi,cao,ma}. 
Initially, inconsistent results on the magnetic structure
were suggested. But, there is now a good agreement between the calculations and 
the experiments in respect of the magnetic and the related lattice structures.
The antiferromagnetic SDW is confirmed to be the lowest energy state, in which 
the orthorhombic structure is stabilized by optimizing the lattice structure 
under the SDW magnetic order \cite{ishibashi}.
However, it is noted that there remains a big puzzle in the magnetic moment 
on an Fe atom in the SDW state.
The moments calculated with the local-density approximation (LDA) and the generalized gradient approximation (GGA)
are $\sim 1.0$ $\mu_{\rm B}$ (LDA) to $\sim 2.0$ $\mu_{\rm B}$ (GGA) \cite{abinitio_matome}, which 
is much larger than experimental data $ 0.2 \sim 0.35 \mu_{\rm B} $ obtained by 
the powder neutron scattering\cite{neutron}, M\"{o}ssbauer effect\cite{mossbauer}, and 
muon spin relaxation \cite{klauss}.
Normally, the moment over-estimation in LDA(GGA) calculation is almost rare. 
This clearly implies that the calculations lack an unknown effect peculiar to 
the compounds or requires a particular correction.    
For example, we immediately notice that spin fluctuations always suppress the moment and 
two-dimensionality due to layered materials somewhat enhances the fluctuation.
However, $S \sim 2$ is too large to be reduced to about $\sim 1/5$ times only by the fluctuations.
Thus, this puzzle may be crucial for elucidating the superconducting mechanism.
In this paper, we report how the puzzle is solved within the framework of LDA + $U$.
We believe that 
the result gives a hint on the superconducting mechanism. 

The Coulomb repulsion effect inside the local Fe $d$-orbit is listed as a missing effect.
The effect is approximately treated within ``LDA + $U$", $U (\equiv U_{\rm eff})$ of which is 
theoretically decomposed into $U_{\rm eff} = U-J$, where $U$ is the on-site Coulomb 
repulsion (Hubbard $U$) and $J$ is the atomic-orbital intra-exchange 
energy (Hund's parameter) in the ``simplified" framework\cite{dudarev}.
We vary the single parameter $U_{\rm eff}=U-J$ from a large positive value to a slightly negative one and 
also calculate electronic structures in another ``separable" framework for $U$ and $J$\cite{liechten} in order to 
identify which parameter is more essential. We find that a slight negative $U_{\rm eff}$ well explains 
the experimental results. 

The on-site Hubbard $U$ employed in the electronic structure calculations is usually a positive value.
The positiveness promotes the localized character of $d$-electrons and enhances 
the magnetic moment in the cases of magnetically ordered compounds. 
Since the parameter U is treated within the mean-field level in the standard LDA+ $U$, more 
sophisticated methods coupled with DMFT and QMC have been also suggested \cite{haule,shorikov}.
However, we point out that one cannot expect a drastic moment reduction 
irrespective of the use of such advanced methods as long as the positive large $U$ is taken into account.
In contrast, we extend the parameter $U_{\rm eff}$ range to a negative one.
Although the negative case is not popular, it occurs in the following two cases.
The Hubbard $U$ itself is negative, and the intra-exchange $J$ is effectively larger than the Hubbard $U$.
The former case has been suggested by many authors 
in various theoretical models as discussed later. 
Besides such explicit reasons, a slightly negative Hubbard $U$ simply may compensate the 
energy calculation errors in the LDA calculations. 
However, there has been just a few works 
taking the negative on-site Hubbard $U$ into account 
within the LDA+ $U$ framework to our knowledge \cite{LDAminusU,persson}.
On the other hand, the latter case \cite{sarma} may be also possible in the present compounds since $J$ is  
suggested to be large \cite{haule2} while $U$ is reported to 
be effectively smaller than our naive expectation \cite{kurmaev,kroll}.
These facts require that one flexibly chooses $U_{eff}$ in the present compounds at least.

\begin{figure}
\includegraphics[scale=0.5]{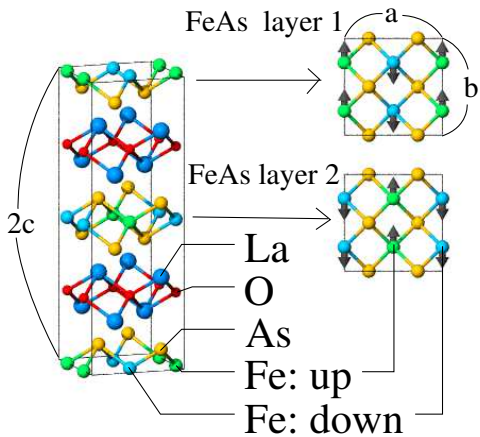}
\caption{The crystal and magnetic structure in the lowest-energy state of the mother compound LaFeAsO 
calculated in the standard LDA ($U_{\rm eff}=0$). 
The arrows on FeAs layers denote the spin direction.
\label{fig:1}}
\end{figure}

In this study, we employ the VASP code\cite{VASP} in which
the projector augmented-wave (PAW) method\cite{PAW,PAWPP} is implemented, and 
the lattice and the electronic structures are optimized by choosing 
GGA + $U$. The stabilities of some different magnetic structures are examined by comparing their 
total energy. The typical lowest-energy state at $U=0$ for the mother compound is shown 
in Fig.~\ref{fig:1}, whose magnetic structure is equivalent with Ref.~\cite{neutron}.
For the mother compound ($x=0$), we actually adopt the La$_4$Fe$_4$As$_4$O$_4$ cell
that have space group symmetry {\it Ibam}, in order to realize
SDW antiferromagnetic state as in Fig.~\ref{fig:1}.
In the case of the doped compound ($x=0.125$), we use the super cell as shown 
in Fig.~\ref{fig:1}, where one of O atoms is replaced by F.
In our calculation, the energy cut-off for the plane-wave set is set to 500 eV, the convergence condition 
for electronic self-consistent loop is less than $10^{-5}$ eV in the total energy difference, and the structure
relaxation loop is repeated until all forces on ions are smaller than 0.02 eV/\AA.
The grids for $k$ points are taken as $6\times 6\times 6$ and $4\times 4 \times 2$
for the undoped and doped compounds, respectively.

\begin{figure}
\includegraphics[scale=0.6]{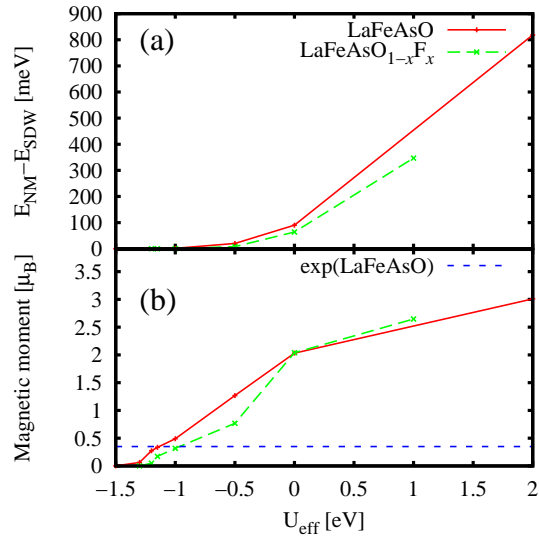}
\caption{ The $U_{\rm eff}$ dependence for the undoped compound (solid curve) LaFeAsO and doped compound (dashed curve)
LaFeAsO$_{1-x}$F$_x$ ($x=0.125$) on
(a) the total energy difference between the SDW (see Fig.~\ref{fig:1}) and the non-magnetic state, and
(b) the SDW magnetic moment on a Fe-site, in which the experimental data (dotted curve)  
taken from Ref.\cite{neutron} is plotted.
\label{fig:2}} 
\end{figure}

Let us show numerical calculation results.
Firstly, we focus on the mother compound LaFeAsO.
Figure \ref{fig:2}(a) shows $U_{\rm eff}$ dependence of 
the energy difference between the non-magnetic and the stable SDW states, and 
Fig.~\ref{fig:2}(b) displays $U_{\rm eff}$ dependence of the magnetic moment in the SDW state.
At $U_{\rm eff}=0$ where several calculations were previously made, the moment 
is $\sim 2.0$ $\mu_{\rm B}$ and the energy difference is $\sim $ 90 meV per LaFeAsO cell.
These values are consistent with the previous results \cite{abinitio_matome,cao,ma,ishibashi},
in which 
the lattice structure is optimized and the 
orthorhombic one is observed to be stable.
For $U_{\rm eff} > 0$, the effect of $U_{\rm eff}$ further develops the moment value.
As expected, the positive $U_{\rm eff}$ stresses the localized character of d-orbital 
electrons, and overemphasizes the inconsistency with the experiments.
On the other hand, the moment decreases with decreasing $U_{\rm eff}$ and coincides with 
the experimental results in a negative $U_{\rm eff}$ region.        
In addition, we note that the energy difference becomes very small around $U_{\rm eff}=-1$, as in Fig.~\ref{fig:2}(a).
This indicates that SDW state becomes not so strongly stable as that at $U_{\rm eff} = 0$.

\begin{figure}
\includegraphics[scale=0.6]{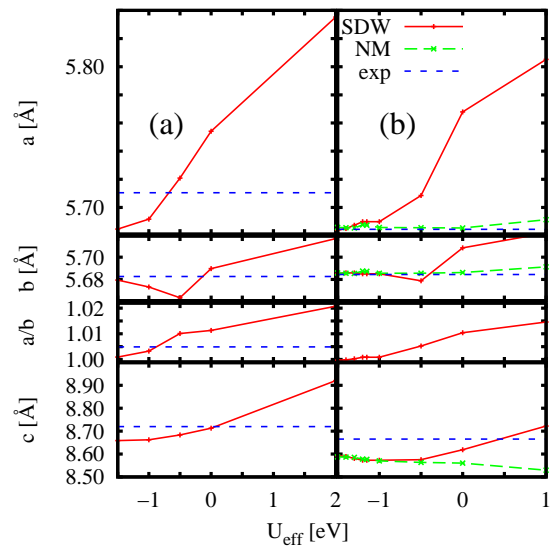}
\caption{ The $U_{\rm eff}$ dependence for (a) the undoped compound LaFeAsO and (b) the doped compound
LaFeAsO$_{1-x}$F$_x$ ($x=0.125$) on
the lattice constants $a$, $b$, $c$, and their ratio ($a/b$) in which the 
experimental data taken from Ref.\cite{nomura} are also plotted. 
The solid curve and dashed curve stand for the SDW and non-magnetic states, respectively.
\label{fig:3}}
\end{figure}

Figure \ref{fig:3}(a) displays $U_{\rm eff}$ dependence of lattice constants, $a$, $b$, and $c$ for 
the mother compound. The experimental values \cite{nomura} are shown by horizontal dashed lines. 
In this paper, $b$ is the direction along which the antiferromagnetic ordering grows.
In the positive $U_{\rm eff}$ range, the increase of $U_{\rm eff}$ brings separations from the 
experimental results for $a$ and $b$, while
they show good agreements with experimental ones in the slightly negative range.

\begin{table}
\caption{The comparison between $U=-1.0(J=0.0)$ and $U=0.0(J=-1.0)$ on the magnetic moment and lattice constants 
for the mother compound in the separable LDA+ $U$. \label{tbl:1}}
\begin{ruledtabular}
\begin{tabular}{ccccrrr}
& & Structure & $\mu$ [$\mu_{\rm B}$]&$a$ [\AA] & $b$ [\AA] & $c$ [\AA] \\
 \hline
\multicolumn{2}{c}{$U-J=-1$}
                & SDW & 0.49 & 5.69166 & 5.67323 & 8.66216 \\
       &        & NM  & ---  & 5.68683 & 5.68703 & 8.64482 \\
$U=-1$ & $J=0$  & SDW & 0.50 & 5.69166 & 5.67323 & 8.66216 \\
       &        & NM  & ---  & 5.68693 & 5.68693 & 8.64482 \\
$U=0$  & $J=1$  & SDW & 0.77 & 5.70995 & 5.65622 & 8.68032 \\
       &        & NM  & ---  & 5.68693 & 5.68693 & 8.64482 \\
\multicolumn{2}{c}{Experiment\footnote{Experimental data are taken from Refs.~\cite{neutron,nomura}}}      
               & SDW & 0.35 & 5.71043 & 5.68262 & 8.71964  
\end{tabular}
\end{ruledtabular}
\end{table}

From the results as seen in Fig.~\ref{fig:2} and \ref{fig:3}(a), it is found that 
the slightly negative $U_{\rm eff}$ well explains the experimental results.
Here, we check which parameter ($U$ or $J$) is more essential in reproducing 
the experimental results by using the separable framework\cite{liechten}.
Table \ref{tbl:1} is a comparison between $U=-1.0$($J=0$) and $J=1.0$($U=0$) for 
the magnetic moment and the lattice constants.
These results indicate that $U=-1.0$ is slightly closer to the experiments but it is not 
a conclusive difference. Moreover, the lattice constants obtained by 
the optimization do not differ significantly in both cases.
We point out that a clear-cut determination is impossible here
within the present framework.

Next, let us turn to the doped compound.
The calculations are made on the supercell as shown in Fig.~1, where 
one of O atoms is replaced by F and $x=0.125$ in LaFeAsO$_{1-x}$F$_{x}$.
At the doping value, the compound 
shows the superconducting ground state.
The dashed lines of Fig.~\ref{fig:2}(b) show $U_{\rm eff}$ dependence for the magnetic moment 
value in the doped compound.
The data calculated at $U_{\rm eff}=0$ still shows the stability of the SDW state,
which is equivalent with the previous literatures \cite{abinitio_matome, ishibashi}.
Then, the energy difference ( $\sim $ 90 meV)  is too large to replace the SDW state by 
the non-magnetic ordered one even in the doped one.
Also, the increase of $U_{\rm eff}$ in the positive range enhances the stability 
of the SDW state, which is inconsistent with the experimental results.
On the other hand, the negative $U_{\rm eff}$ diminishes the moment resulting in 
the disappearance of the SDW, 
and the system recovers to the tetragonal lattice structure by the optimization.
The behavior of the lattice constants is also 
similar to the undoped case as seen in Fig.~\ref{fig:3}(b).  
At the negative range, the calculated values except for $c$ 
show good agreements with experimental ones.

\begin{figure}
\includegraphics[scale=0.5]{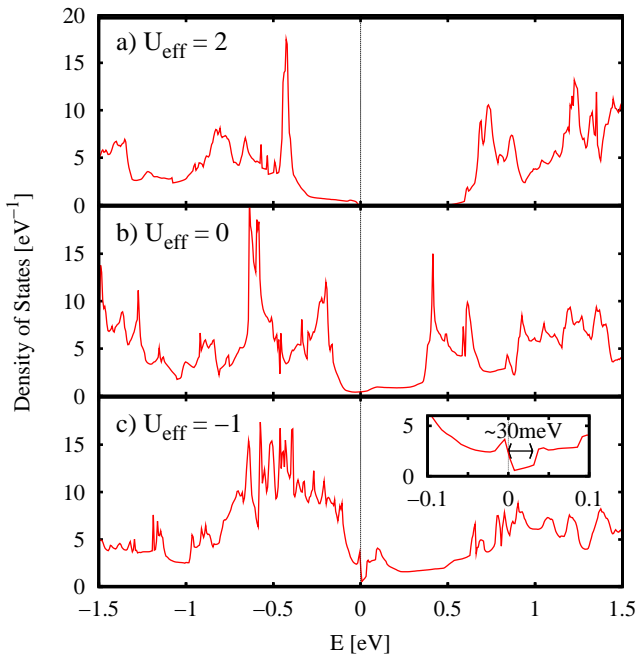}
\caption{The electronic density of states DOS for (a) $U_{\rm eff}=2$, (b) $U_{\rm eff}=0$, and 
(c) $U_{\rm eff}=-1$ in the undoped compound.
$E=0$ corresponds to the Fermi energy.
\label{fig:4}
}
\end{figure}

Figure \ref{fig:4} shows $U_{\rm eff}$ dependent features of the electronic DOS around the Fermi level in 
the case of the mother compound.
Three cases, namely, $U_{\rm eff}$ positive, zero, and negative ones are given 
in Figs.~\ref{fig:4}(a), 4(b), and 4(c), respectively.
At $U_{\rm eff}=0$, the gap like DOS suppression whose width is over $\sim 300$ meV opens, and 
the positive $U_{\rm eff}$ simply expands the width, which results in 
much overestimation compared to the pseudo-gap value reported by 
recent experiments \cite{pseudogap,dong}.  
On the other hand, the slightly negative $U_{\rm eff}$ gives the gap value much closer to 
the experimental ones ($\sim$ 30 to 100 meV) as shown in Fig.~\ref{fig:4}(c).

Finally, let us discuss the mechanism of the negative $U_{\rm eff}$.
Firstly, we give a summary of the studies on the negative Hubbard $U$.
Many authors have suggested their ideas on the 
negative Hubbard $U$, which are summarized in terms 
of theoretical target materials as follows,
i) several inorganic compounds forming the charge density wave,  
ii) amorphous semiconductors like chalcogenide glasses, iii) conducting 
polymers, and iv) heavy-fermion systems.
These were intensively debated in old literatures related to the local electron pairing \cite{micnas}.
On the other hand, more advanced studies inspired by the discovery of cuprate High-$T_{\rm c}$ superconductors
have been published on several metal oxides.
A famous idea originates from the charge disproportionation on the cation sites, which 
has been mainly discussed in doped superconductors based on BaBiO$_3$, in which 
CDW occurs close to the emergence of its superconductivity \cite{hase}.
In addition, the overscreening effects on the Coulomb interaction 
coupled with the strong correlation (due to the low carieer density) have been proposed 
in the context of high-$T_{\rm c}$ superconductivity 
mechanism \cite{tachiki}. If the overscreening, i.e., the attractive interaction effectively works 
between two electrons, then unusual softening of electronically coupled 
longitudinal optical (LO) phonon compared to transverse optical (TO) one 
has been predicted. In fact, there are several reports which confirm the LO-TO frequency 
inversion due to the drastic LO softening in high-$T_{\rm c}$ superconductor as well as other various metal oxides \cite{graf} .
Thus, the negative Hubbard $U$ is now not rare. 
But, since the effect is counterintuitive, more experimental and theoretical tasks are required.  
On the other hand, the idea that $J$ is relatively effective in fixing $U_{eff}$ may be very simple and acceptable.
In this iron based superconductors, the Hubbard $U$ may be rather small because Fe related five bands are 
entangled within their wide band-width. In fact, there are some experimental 
reports which support it \cite{kurmaev,kroll}.

In conclusion, we calculated electronic structure of iron based typical 
superconducting compound LaFeAsO$_{x}$F$_{1-x}$ using the 
framework LDA+$U$ with expanding the range of $U (\equiv U_{\rm eff}= U-J)$ from the positive to 
the slightly negative range.
Consequently, we found that the calculated magnetic moment, 
the lattice constants, and the pseudo-gap feature shows good agreements with the experimental results in 
the negative $U_{\rm eff}$ range. We discussed some negative origin of $U_{\rm eff}$ in 
the superconductor.

The authors wish to thank S.~Shamoto, A.~Baron, M.~Okumura, T.~Koyama, 
A.~Fujiwara, T.~Maehira, and N.~Hamada for illuminating discussion. 
The work was partially supported by the
Priority Area ``Physics of new quantum phases in superclean materials''
(Grant No.~18043022) from the Ministry of Education, Culture, Sports,
Science and Technology of Japan, and also a JAEA internal project represented 
by M.Kato. One of authors (M.M.) is supported by
JSPS Core-to-Core Program-Strategic Research Networks, ``Nanoscience and
Engineering in Superconductivity (NES)''. 


\end{document}